\documentclass[11pt]{article}

\usepackage{amsmath}
\usepackage{amsthm}
\usepackage{amssymb}
\usepackage{mathtools}
\usepackage{hyperref}

\newcommand{\BA}{\begin{array}}
\newcommand{\EA}{\end{array}}
\newcommand{\BE}{\begin{enumerate}}
\newcommand{\EE}{\end{enumerate}}
\newcommand{\BI}{\begin{itemize}}
\newcommand{\EI}{\end{itemize}}

\newcommand{\I}{\item}

\def\BEQ#1\EEQ{\begin{align*}#1\end{align*}}
\def\BNEQ#1\ENEQ{\begin{align}#1\end{align}}

\newcommand{\BPF}{\begin{proof}}
\newcommand{\EPF}{\end{proof}}

\newcommand{\B}[1]{\left( {#1} \right)}
\newcommand{\F}[2]{{\frac{#1}{#2}}}
\newcommand{\R}[1]{{\frac{1}{#1}}}
\newcommand{\BF}[2]{\B{\F{#1}{#2}}}
\newcommand{\set}[1]{\left\{ #1 \right\}}
\newcommand{\midline}[2]{{#1} \, \middle| \, {#2}}
\newcommand{\prob}[1]{\text{\tt Pr}\left[ #1 \right]}
\newcommand{\expct}[1]{\text{\tt E}\left[ #1 \right]}

\newcommand{\stset}[2]{\set{\midline{#1}{#2}}}

\newcommand{\Real}{\mathbb{R}}

\usepackage[margin=0.7in]{geometry}
\usepackage{nicefrac}
\usepackage{authblk}
\usepackage{algorithm}
\usepackage[noend]{algpseudocode}

\title{Parallel Linear Search with no Coordination\\ for a Randomly Placed Treasure\footnote{This work has received funding from the European Research Council (ERC) under the European Union's Horizon 2020 research and innovation programme (grant 
agreement No 648032).}}

\author[1]{Amos Korman}
\author[2]{Yoav Rodeh}
\affil[1]{CNRS and University Paris Diderot.\\
  Paris, France\\
  \texttt{amos.korman@irif.fr}}
\affil[2]{Weizmann Institute of Science\\
  Rehovot, Israel\\
  \texttt{yoav.rodeh@gmail.com}}

\theoremstyle{definition}
\newtheorem{definition}{Definition}

\theoremstyle{plain}
\newtheorem{claim}{Claim}
\newtheorem{theorem}{Theorem}
\newtheorem{corollary}{Corollary}
\newtheorem{lemma}{Lemma}

\newcommand{\speedup}{\textit{speedup}}
\newcommand{\zoom}[3]{{#1}_{\overrightarrow{#2,#3}}}
\newcommand{\column}{C}
\newcommand{\ceil}[1]{{\left\lceil {#1} \right\rceil}}
\newcommand{\OPT}[1]{{\text{OPT}_{#1}}}
\newcommand{\cont}{{\cal F}}
\newcommand{\valid}{{\cal V}}

\begin{document}

\maketitle

\begin{abstract}

In STOC'16, Fraigniaud et al.\  consider the problem of finding a treasure hidden in one of many boxes that are ordered by importance. That is, if a treasure is in a more important box, then one would like to find it faster. Assuming there are many searchers, the authors suggest that using an algorithm that requires no coordination between searchers can be highly beneficial. Indeed,  besides saving the need for a  communication and coordination mechanism, such algorithms enjoy inherent robustness. The authors proceed to solve this linear search problem in the case of countably many boxes and an adversary placed treasure, and prove that the best speed-up possible by $k$ non-coordinating searchers is precisely $\frac{k}{4}(1+1/k)^2$. In particular, this means that asymptotically, the speed-up is four times worse compared to the case of full coordination.

We suggest a natural variant of the problem, where the treasure is placed uniformly at random in one of a finite, large, number of boxes. We devise non-coordinating algorithms that achieve a speed-up of $6/5$ for two searchers, a speed-up of $3/2$ for three searchers, and in general, a speed-up of $k(k+1)/(3k-1)$ for any $k \geq 1$ searchers. Thus, as $k$ grows to infinity, the speed-up approaches three times worse compared to the case of full coordination. Moreover, these bounds are tight in a strong sense as no non-coordinating search algorithm for $k$ searchers can achieve better speed-ups. We also devise non-coordinating algorithms that use only logarithmic memory in the size of the search domain, and yet, asymptotically,  achieve the optimal speed-up. Finally, we note that all our algorithms are extremely simple and hence applicable.
\end{abstract}

\pagenumbering{gobble}
\clearpage
\pagenumbering{arabic}

\section{Introduction}

Parallel algorithms are often evaluated by their {\em speed-up}, which measures how much faster the parallel algorithm with $k$ processors runs in comparison to the best running time that a single processor can achieve. In \cite{fraigniaud_parallel_2015}, the authors focused on  parallel {\em non-coordinating} algorithms.  In such an algorithm, all processors operate independently, by executing the same protocol, differing only in the outcome of the flips of their random coins. A canonical example is the case of multiple random walkers that search in a graph \cite{alon_many_2008}. 
Although most problems cannot be efficiently parallelized without coordination, when such parallelization can be achieved,  the benefit can potentially be high. Particularly, as argued in \cite{fraigniaud_parallel_2015}, the gain can be not only in terms of saving in communication and overhead in computation, but also in terms of robustness.

One class of fundamental problems that enjoys a large non-coordination ratio in some circumstances, is the class of search problems over totally ordered sets.  The objective of such \emph{linear search} problems is to find a solution among a set of candidate solutions that are linearly ordered according to their quality. For instance, searching for a proper divisor of a random integer $n$ is an illustration of linear search. In this case, enumerating the candidate divisors in increasing order, and checking them one after another, is the typical strategy to solve the problem, as the probability that $n$ is divisible by a given prime is inversely proportional to this prime. Similarly, in cryptography, an exhaustive search attack is better proceeded by systematically checking smaller keys than longer ones, since
checking a key typically requires exponential time in the key size. In general, linear search appears in contexts in which the search space can be ordered in a way such that, given that the previous trials failed, the next candidate according to the order is either the most likely to be valid, or most preferable, or the easiest to check.

One basic linear search problem is the {\em treasure-hunt} problem. There is one treasure hidden in one box $B_x$ out of a linearly ordered set of boxes $\{B_i\}$. Searchers are unaware of the index $x$ and their goal is to find the treasure as fast as possible. At each time step, a protocol executed by a searcher has the ability to peek into one box and see whether the treasure is present or not in that box. The protocol is  terminated once one of the searchers finds the treasure. Boxes are listed in an order that reflects the importance of finding the treasure in a given box. 
That is, finding the treasure hidden in $B_i$ for small values of~$i$ is more urgent than for large values of~$i$. In the case of a solo searcher, the best strategy is to open the boxes in increasing order. In this case, the searcher will find the treasure in $x$ time. Hence, for a given  algorithm $A$, we  measure the {\em speed-up} function of~$k$ searchers with respect to $x$ as:
\BEQ
\speedup_{A}(k,x)=\frac{x}{\expct{\text{time to find $x$ with $k$ searchers running  $A$}}}
\EEQ
where the expectation is taken with respect to the randomness of the searchers. Note that if coordination is allowed,  one could simply divide the set of boxes evenly between the searchers leading to a speed-up of $k$. However, as mentioned in \cite{fraigniaud_parallel_2015}, as simple as this algorithm is, it is very sensitive to faults of all sorts. For example, if one searcher crashes at some point in time during the execution then the searchers may completely miss the treasure, unless the protocol employs some mechanism for detecting such faults. This observation motivates the study of non-coordinating algorithms. 

In \cite{fraigniaud_parallel_2015} the authors considered the case in which an adversary places the treasure in one of infinitely countable boxes. 
In this case, for any non-coordinating algorithm, the adversary can always find a box for which the speed-up would be at most 1. Therefore, 
 the speed-up of an algorithm was defined as $\limsup_{x \rightarrow \infty}\speedup_{A}(k,x)$.  
They then present Algorithm \ref{alg:stoc} that is very simple and achieves optimal performance, 
a speed-up of $\frac{k}{4}(1+1/k)^2$ for each $k$. Specifically, the speed-up is
$9/8$ for two searchers, $4/3$ for three searchers, and roughly ${k}/{4}$ for $k$ searchers as $k$ grows larger.
It is shown that no non-coordinating search algorithm can achieve better speed-ups. 
\begin{algorithm}
\caption{The optimal algorithm in \cite{fraigniaud_parallel_2015} for the adversary placed treasure.}\label{alg:stoc}
\begin{algorithmic} 
\For{$i=1$ to $\infty$}
  \State check a uniformly chosen box of those unchecked in $\set{1, \ldots, i*(k+1)}$
  \State check another uniformly chosen box of those unchecked in the same range
\EndFor
\end{algorithmic}
\end{algorithm}

In this paper, we expand on the techniques and results of \cite{fraigniaud_parallel_2015}, and solve a natural variant of the treasure-hunt problem. In our scenario, the treasure is placed {\em at random} in one of $M$ boxes. 
We therefore define the speed-up of an algorithm as:
\BEQ
\speedup_{A}(k)=\expct{\speedup_{A}(k,x)}
\EEQ
where the expectation is taken with respect to the placement of $x$.
The speed-up of $k$ agents on $M$ boxes is defined as the speed-up of the best algorithm for $M$ boxes. That is:
\BEQ
\speedup_M(k)=\sup \{\speedup_{A}(k)\mid \text{$A$ works on $M$ boxes}\}.
\EEQ
Finally, the general speed-up of $k$ agents is defined as the speed-up of the best algorithm for $M$ boxes, where $M$ is taken to infinity. That is:
\BEQ
 \speedup(k)= \liminf_{M\rightarrow \infty} \speedup_M(k).
\EEQ

\subsection{Our Results}
We suggest a general framework to tackle the randomly placed treasure hunt problem and solve it completely for the case of a treasure uniformly placed in one of finitely many boxes.

Specifically, we prove that the optimal speed-up for non-coordinating algorithms running with $k\geq 1$ searchers is:
\BEQ
\speedup(k)=\F{k(k+1)}{3k-1}
\EEQ
This means that the best speed-up that can be achieved is ${6}/{5}$ for two searchers, ${3}/{2}$ for three searchers, and roughly $k/3$ for $k$ searchers, as $k$ grows larger. We stress that these bounds are tight in a strong sense as no non-coordinating search algorithm for $k$ searchers can achieve better speed-ups, for any $k$. 

Our results indicate that the  best possible speed-up for the random setting is strictly higher than 
the best possible speed-up for the adversarial setting. Specifically, as the number of searchers grows, the multiplicative gap between the settings approaches $4/3$.

The non-coordinating algorithm achieving\footnote{Note that this algorithm achieves the bound in an asymptotic manner, meaning that as the number of boxes goes to infinity, the speed-up of the algorithm approaches the desired bound. For small number of boxes there might be better algorithms.} the aforementioned ratio for the random setting with $k$ searchers is so simple that it can be described in just a few lines (Algorithm \ref{alg:new}).
Note that for each $k$ and $M$ the algorithm is different.
\begin{algorithm} 
\caption{The asymptotically optimal algorithm.}\label{alg:new}
\begin{algorithmic} 
\For{$i=1$ to $M/k$}
   \State check a uniformly chosen box from those unchecked in $\set{1, \ldots, i*k}$
\EndFor
\While{there are unchecked boxes}
	\State check a uniformly chosen box from those unchecked
\EndWhile
\end{algorithmic}
\end{algorithm}
It is interesting to note that this algorithm finds a way to balance between two opposing forces: it gives higher probability of quickly opening the more important boxes, and yet tries to minimize as much as possible the overlap between different agents.

Despite the simplicity of Algorithm \ref{alg:new}, finding it was not an easy task. Indeed, there are many possible algorithms that achieve an $O(k)$ speed-up, but coming up with an algorithm that has an exact optimal bound is not trivial. This is especially challenging as every slight change in the algorithm may significantly complicate its analysis. 

Algorithm \ref{alg:new} achieves optimal speed-up for each $k$ but requires $\Omega(M)$ bits of memory to store the elements that have already been checked. To circumvent this, we suggest and analyse a natural variant of this algorithm, which works in exactly the same fashion, except that it does not remember the boxes it already checked. It therefore requires only $O(\log M)$ bits of memory, and yet its speed-up is $k/3$, which is practically the same as that of the optimal algorithm for large $k$.

Concerning the technical aspects, 
in a similar fashion to \cite{fraigniaud_parallel_2015}, we first observe that the crucial aspects of non-coordinating algorithms can be represented by matrices and then approximate matrices by continuous functions.
To show the upper bound on the speed-up we had to upper bound the integral of functions satisfying certain restrictions. As in \cite{fraigniaud_parallel_2015}, the biggest technical challenges lie in computing such integrals. This task turned out to be more difficult than in the adversarial case of \cite{fraigniaud_parallel_2015}, and we were therefore unable to directly compute these integrals. Instead, to overcome this, we needed two new tools: We extend a lemma of \cite{fraigniaud_parallel_2015} to work on unbounded functions,  and introduce the intuitive notion of the ``zooming'' of a function. Using this two new results,  we can reduce the problem to solving simpler integrals.  
We then turn to analyze Algorithm \ref{alg:new} to lower bound the speed-up for $k$ agents. 
One of the problems arising in the analysis of Algorithm \ref{alg:new}which did not occur in the limsup adversarial setting of \cite{fraigniaud_parallel_2015}, is that the first boxes cannot be ignored, as they contribute significantly to the average case analysis. To overcome this we prove a small, yet useful, result regarding the Gamma function.

As mentioned, we fully characterize the case of a treasure placed uniformly at random. We note, however, that 
the general framework of analysis presented in this paper could be useful in order to derive speed-up bounds with respect to other distributions as well. Handling other distributions, however, would require solving the corresponding particular integrals.  

To sum up, our upper bound on the speed-up of  non-coordination algorithms  implies  that there is a price to be paid for the absence of coordination, which is asymptotically a factor of three away from an ideal optimal algorithm that  performs with perfect coordination. On the other hand, this price is actually  reasonably low, and so in faulty contexts in which coordination between the searchers may yield severe overheads, it might well be worth giving up on coordination, and simply run our non-coordinating algorithm.

\subsection{Related Work}

Most of the parallel search literature deals with mobile agents searching graphs of different topologies. One example is search by multiple random walkers. In a series of papers \cite{alon_many_2008, cooper_multiple_2009, elsasser_tight_2011,efremenko_how_2009} several  results regarding hitting time, cover time and mixing times are established, such as a linear speed-up for several graph families including expanders and random graphs.

Another classical example of mobile search is the linear search problem \cite{beck_linear_1963, bellman_optimal_1963} (and later \cite{gal_minimax_1974}), where a treasure is hidden on the real line according to a known probability distribution. A searcher starts from the origin and wishes to discover the treasure in minimal expected time.
This problem was reintroduced by computer scientists as the {\em cow-path} problem \cite{baezayates_searching_1993,kao_searching_1993}.
There, similarly to this paper, the time complexity is measured as a function of the distance of the treasure from the starting point.

Motivated by applications to central search foraging by desert ants, the authors in \cite{feinerman_memory_2012,feinerman_collaborative_2012} considered the ANTS problem, a variant of the cow-path problem on the grid, and showed that a speed-up of $O(k)$ can be achieved with $k$ independent searchers. Several variants of the cow-path problem and the ANTS problem were studied in  \cite{demaine_online_2006, emek_how_2015, emek_solving_2014,karp_search_1986, langner_fault-tolerant_2014, lenzen_trade-offs_2014}.

The major difference between our setting and the mobile agent setting, is that we allow ``random access'' to the different boxes. That is, our searcher can jump between different boxes at no cost, unlike the case of mobile agents on a graph which can only move from a vertex to a neighboring vertex in one time step.

\section{Preliminaries}

In this section we go over the setup of \cite{fraigniaud_parallel_2015} with some minor changes that reflect our changed definition of speed-up. We then proceed to introduce some new tools to help us deal with the case of randomly placed treasure.

\subsection{Setup}
Our universe contains boxes indexed by $i \in \set{1 \ldots M}$, and a treasure that is placed uniformly at random in one of them. At each time step, a searcher can peek into exactly one box. There are $k$ searchers and they are all completely identical in that they have the same algorithm, yet their randomness is independent. They do not communicate at all, and their mutual purpose is to maximize the speed-up by minimizing the expected time until the first one of them finds the treasure. In our technical discussion it will be often easier to work with the {\em inverse} of the speed-up. Specifically, for an algorithm $A$ and an index $x$, define:
\BEQ
\theta_A(k,x) = \R{x} \expct{\text{time to find $x$ with $k$ searchers running $A$}} 
\EEQ
where the expectation is taken over the randomness of the searchers. Taking expectation over the placement of $x$, we define:
\BEQ
\theta_A(k) = \expct{ \theta_A(k,x) } 
\EEQ
So, an algorithm with $\theta(3) = 1/2$ means that running this algorithm on three searchers will result in an expected running time that is twice as fast as the trivial one-searcher algorithm.

\subsection{From Algorithms to Matrices}

Given an algorithm, we define the matrix $N$, where
$N(x,t)$ is the probability that the algorithm executed with just one searcher has {\em not} visited box $x$ up to (and including) step~$t$. 
For example, the following matrix corresponds to Algorithm \ref{alg:new}, where $k=2$ and $M=6$ (note that we stop after time $t=6$ since afterwards, the matrix is all 0's).
\newcommand{\NF}[2]{{\nicefrac{#1}{#2}}}
\newcommand{\OT}{\nicefrac{1}{3}}
\newcommand{\TT}{\nicefrac{2}{3}}
\BEQ
\BA{c|cccccccc}
_{x\downarrow}^{t\rightarrow} & 0 & 1 & 2 & 3 & 4 & 5 & 6 & \ldots\\
\hline
1 & 1  & \NF12 & \NF13 & \NF14    & \NF16  & \NF{1}{12} & 0  & \\
2 & 1  & \NF12 & \NF13 & \NF14    & \NF16  & \NF{1}{12} & 0  & \ldots \\
3 & 1  & 1     & \NF23 & \NF12    & \NF13  & \NF16      & 0  & \\
4 & 1  & 1     & \NF23 & \NF12    & \NF13  & \NF16      & 0  & \ldots \\	
5 & 1  & 1     & 1     & \NF34    & \NF12  & \NF14      & 0  & \\
6 & 1  & 1     & 1     & \NF34    & \NF12  & \NF14      & 0  & \ldots \\
\EA
\EEQ
Some observations:
\BE
\I
The value of $N(x,t)$ is $N(x,t-1)$ multiplied by the probability that $x$ will not be checked at time $t$. This is how we arrived at the matrix above.
\I
The sum of row $x$ is the expected time until the algorithm peeks into box $x$.
Indeed, let $I_{x,t}$ denote the indicator random variable that is 1 iff $t <$ the visit time of $x$.
The sum of these over $t$ is the visit time. Also, 
$\Pr[I_{x,t} = 1] = N(x,t)$,
so we get the result by linearity of expectation. This means:
\BEQ
\theta_N(1,x) = \R{x} \sum_{t=0}^\infty N(x,t)
\EEQ
\I
Given the matrix $N$ for one searcher, what would be the $N$ matrix for $k$ searchers? The probability of $x$ not being looked into up to step $t$ is the probability that all $k$ searchers didn't peek into it, which is $N(x,t)^k$. So by the same reasoning as the last point, we get:
\BNEQ\label{eq:ifactor}
\theta_N(k,x) = \R{x} \sum_{t=0}^\infty N(x,t)^k 
\ENEQ
\I
Since $1-N(x,t)$ is the probability that box $x$ was peeked into by step $t$, summing these numbers over column $t$, we get the expected number of boxes checked by this time,
which is of course at most $t$.
Denoting:
\BEQ
\column_N(t) = \sum_{x=1}^M 1-N(x,t)
\EEQ
The {\em column requirement} is that for all $t$, $\column_N(t) \leq t$.
The matrix resulting from an algorithm will always satisfy the column requirement.
\EE

\subsection{One Searcher} \label{sct:one}

A simple and intuitive result is when considering just one agent. We show it here for completeness and also as intuition, as it is a very simple version of the proof for the general~case.
\begin{theorem} \label{oneAgent}
$\speedup(1) = 1$.
\end{theorem}
\BPF
The algorithm $A$ that opens one box after another will take time $x$ to find the treasure if it is in box $x$, and so:
$\speedup_A(1,x) = x/x = 1$, meaning that $\speedup_A(1) = 1$. 
So, $\speedup(1) \geq 1$.

On the other hand, take some algorithm and denote its matrix by $N$.
\BNEQ \label{eq:oneSearcher}
\theta_N(1) 
& = 
\R{M} \sum_{x=1}^M \R{x} \sum_{t=0}^\infty N(x,t) 
= 
\R{M} \sum_{t=0}^\infty \sum_{x=1}^M \R{x} N(x,t) 
\ENEQ
We aim to show that $\theta_N(1)\geq 1$. For this purpose, we actually show that this inequality holds for every $N$ taking values in $[0,1]$ satisfying  the column requirement.  That is, we know that for every $t$,
\BEQ
C_N(t) = \sum_{x=1}^M 1-N(x,t) \leq t 
\EEQ
Wanting to minimize \eqref{eq:oneSearcher} under this restriction means that we can deal with each column separately. 
So the question is, given some integer $t$, what are the values of $f_1, \ldots, f_M \in [0,1]$ that minimize $\sum_{x=1}^M f_x/x$ under the restriction that $\sum_{x=1}^M (1-f_x) \leq t$. The answer is $f_x = 0$ for all $x\leq t$ and $1$ for the others. Otherwise there is some $y>x$, s.t.\ one can decrease $f_x$ a little and increase $f_y$ by the same amount. This will result in a smaller sum without violating the constraint.

The resulting optimal matrix is therefore $N(x,t) = 0$ for all $x \leq t$ and $1$ otherwise. This is exactly the matrix of the trivial algorithm, and so the optimal speedup is at most $1$.
\EPF

\subsection{From Matrices to Functions}

It will be much easier to work with a continuous version of our problem.
For an interval $X \subseteq \Real^+$ denote:
$\cont(X) = \stset{N : X \times [0,\infty] \rightarrow [0,1]}{N \text{ is continuous}}$.
For an $N \in \cont(X)$, we say that $N$ satisfies the column requirements if for all $t$:
\BEQ
\column_N(t) = \int_X 1-N(x,t) dx \leq t 
\EEQ
We call such a function valid, and denote by $\valid(X)$ the set of all such valid functions.
We define:
\BEQ
\theta_N(k) = \R{|X|} \int_X { \int_0^\infty \R{x} N(x,t)^k dt } dx
= \R{|X|} \int_0^\infty { \int_X \R{x} N(x,t)^k dx } dt
\EEQ
where we can change the integral because of Tonelli's theorem.
The following claim shows a connection between matrices and functions:
\begin{claim} \label{clm:Mat2Func}
For any algorithm $A$ that works on $M$ boxes, there is a function $N \in \valid([1,M+1])$ such that for all $k$, $\theta_N(k) \leq \theta_A(k)$.
\end{claim}
The proof is in Appendix \ref{apx:matToFunc}, where we choose $N$ to be a continuous version of the step function derived from $A$'s matrix.

Claim \ref{clm:Mat2Func} shows that upper bounding the speed-up of all functions in $\valid([1,M+1])$ will upper bound the speed-up of all algorithms on $M$ boxes.

\section{Zooming in (or out) a Function}

We introduce a new and important tool that will be used a couple of times in what proceeds.
If we have some algorithm that works for $M$ boxes, we would like to be able to use it on $2M$ boxes. Slowing it down by a factor of two, and using the extra probability to cover the extra boxes can do the trick. How to do this for algorithms is not clear, but for continuous functions it is simple and elegant.
\begin{definition}
Given some $N \in \cont(X)$ and some $a,b > 0$, we define the zooming of $N$ by $(a,b)$ as: 
\BEQ
\zoom{N}{a}{b}(x,t) = N(x/a, t/b)
\EEQ
Where $\zoom{N}{a}{b}(x,t) \in \cont(aX)$.
\end{definition}
The intuitive meaning of it is that we expanded the algorithm to work on a domain of size $a$ times the original one, and slowed it down by a factor of $b$. 
What happens to the column integrals and to the speed-up?

\begin{claim} \label{clm:zoom}
For $N \in \cont(X)$ and $a,b > 0$, we have 
$
\theta_{\zoom{N}{a}{b}}(k) = \F{b}{a} \theta_N(k)
$
and for all values $t$, 
$
\column_{\zoom{N}{a}{b}}(t) = a \column_N\BF{t}{b}
$.
\end{claim}
\BPF
The inverse speed-up:
\BEQ 
\theta_{\zoom{N}{a}{b}}(k)
& =  
\R{|aX|}\int_{aX} \B{\int_0^\infty \R{x} \zoom{N}{a}{b}(x,t)^k dt} dx 
=
\R{a|X|}\int_{aX} \B{\int_0^\infty \R{x} N(x/a, t/b)^k dt} dx 
\\ & =
\R{a|X|}\int_{aX} \B{ b \int_0^\infty \R{x} N(x/a, t)^k dt } dx
= 
\R{a|X|}\int_{X} \B{ ba \int_0^\infty \R{ax} N(x, t)^k dt } dx
\\ & = 
\F{b}{a} \cdot \R{|X|} \int_{X} \B{ \int_0^\infty \R{x} N(x, t)^k dt } dx
= \F{b}{a} \theta_N(k)
\EEQ
And the column integrals:
\BEQ
C_{\zoom{N}{a}{b}}(t)  = 
\int_{aX} 1-\zoom{N}{a}{b}(x,t)dx = 
\int_{aX} 1-N(x/a, t/b)dx 
 = 
a \int_X 1-N(x, t/b)dx = a C_N\BF{t}{b}
\EEQ
\EPF
\begin{corollary} \label{cr:doubleZoom}
For $N \in \valid(X)$ and any $a > 0$, we have that $\zoom{N}{a}{a} \in \valid(aX)$ and has exactly the same speed-up as $N$.
\end{corollary}
\BPF
By Claim \ref{clm:zoom} and the fact that $N$ satisfies the column requirements, $\column_{\zoom{N}{a}{a}}(t) = a\column_{N}(t/a) \leq a (t/a) = t$. By the same claim we see that the speedup remains unchanged.
\EPF

\section{Upper Bound}

\begin{theorem} \label{thm:upper}
For all $k \geq 1$:
\BEQ
\speedup(k) \leq \F{k(k+1)}{3k - 1}
\EEQ
\end{theorem}
To prove the theorem, we show that for every $\delta > 0$, for large enough $M$, any algorithm $A$ working on $M$ boxes  satisfies:
\BEQ
\speedup_A(k) \leq (1+\delta) \F{k(k+1)}{3k - 1}
\EEQ
The case $k=1$ was dealt with in Section \ref{sct:one}.
Claim \ref{clm:Mat2Func} says that if we prove this for all $N \in \valid([1,M+1])$ then we have proved it on all algorithms designed for $M$ boxes.  Corollary \ref{cr:doubleZoom} then says that by zooming $N$ by $M+1$, proving the following claim  will prove Theorem~\ref{thm:upper}:
\begin{claim} \label{clm:upper}
There is some real function $f$, s.t. $\lim_{\epsilon \rightarrow 0} f(\epsilon) = 1$, and 
every $N \in \valid([\epsilon, 1])$ satisfies:
\BEQ
\theta_N(k) \geq f(\epsilon) \cdot \F{3k-1}{k(k+1)}
\EEQ
\end{claim}
We use a lemma\footnote{It is slightly changed here: the upper bound of the domain is 1 and not infinity. This changes nothing in the original proof and actually simplifies the statement of the lemma, as boundedness and integrability of $a(x)$ are immediate. We also change the requirement to inequality, and this again, changes nothing, as when the optimal solution is not 0, it will clearly take the equality.} of \cite{fraigniaud_parallel_2015}: 
\begin{lemma} \label{lm:oldLemma}
Fix $\epsilon \geq 0$ and $T\geq0$. For continuous functions $a: [\epsilon,1] \rightarrow \Real^+$ and $f : [\epsilon,1] \rightarrow [0,1]$ where $\int_\epsilon^1 1 - f(x) dx \leq T$, 
the minimum of $\int_\epsilon^1 a(x) f(x)^k dx$, ranging over all possible $f$'s is achieved when
\BEQ
f(x) = \min\B{1, \alpha a(x)^{-\R{k-1}}}
\EEQ
where $\alpha$ is a function of $a(\cdot)$ and $T$, and independent of $x$. 
\end{lemma}
This lemma is exactly what we need to prove Claim \ref{clm:upper}. Informally, considering $N \in \valid([\epsilon, 1])$, the column requirement is the condition on $f$ in the lemma, just setting $T = t$, and by plugging in $a(x) = 1/x$, we get the form of the optimal function for each of the columns. For each $t$, we then find the maximal $\alpha$ solving
$\int_\epsilon^1 1 - \min(1, \alpha x^{1/(k-1)})dx \leq t$, 
and thus get the optimal function $N$ for a specific $\epsilon$. We can then calculate $\theta_N(k)$ and take the limit as $\epsilon$ goes to 0 (that is, $M$ goes to infinity) to get our result.

The only problem with this approach is that the integrals become difficult to solve when we start them at some arbitrary $\epsilon$. We therefore take $\epsilon=0$. Then, as we shall see, the integrals are pretty straightforward. Our plan is therefore as follows:

\BE
\I
Extend Lemma \ref{lm:oldLemma}. 
This is because it doesn't work now, as $a(x) = 1/x$ is not defined on~$0$, and cannot be extended in a continuous way to it.
\I
Use the extended lemma to find the exact optimal function in the case of $\epsilon=0$, and calculate its speed-up.
\I
Use the result to come back to $\epsilon > 0$ with asymptotically insignificant change in speed-up.
\EE

This is the extended lemma:
\begin{lemma} \label{lm:newLemma}
Fix $T \geq 0$. For continuous functions $a: (0,1] \rightarrow \Real^+$ and $f : (0,1] \rightarrow [0,1]$ where $\int_0^1 a(x)^{-1/(k-1)} dx$ is defined and $\int_0^1 1 - f(x) dx \leq T$, 
the minimum of $\int_0^1 a(x) f(x)^k dx$, ranging over all possible $f$'s is achieved when:
\BEQ
f(x) = \min\B{1, \alpha a(x)^{-\R{k-1}}}
\EEQ
Where $\alpha$ is a function of $a(\cdot)$ and $T$, and independent of $x$. 
\end{lemma}
It is proved in Appendix \ref{apx:mainLemma} by using the original lemma on subsets $[s,1]$, taking the limit of the resulting functions, and then using the dominated convergence theorem.

As we can see the condition on $a$ is a bit bizarre, but it proves what we need:
\begin{corollary} \label{cr:main}
Among all continuous functions $f : (0,1] \rightarrow [0,1]$ satisfying
$\int_0^1 1-f(x)dx \leq T$, the one that minimizes $\int_0^1 \R{x} f(x)^k dx$ is
$
f(x) = \min\B{1, \alpha x^\R{k-1}}
$,
where $\alpha$ is a function of~$T$.
\end{corollary}
\BPF
Set $a(x) = 1/x$. To use Lemma \ref{lm:newLemma}, we check that 
$
\int_0^1 a(x)^{-\R{k-1}} dx = \int_0^1 x^\R{k-1} dx < \infty
$,
and this is true for all $k \geq 1$.
\EPF

\subsection{The Optimal Function $\OPT{k}$}

\begin{figure} 
\center
\includegraphics[scale=0.3]{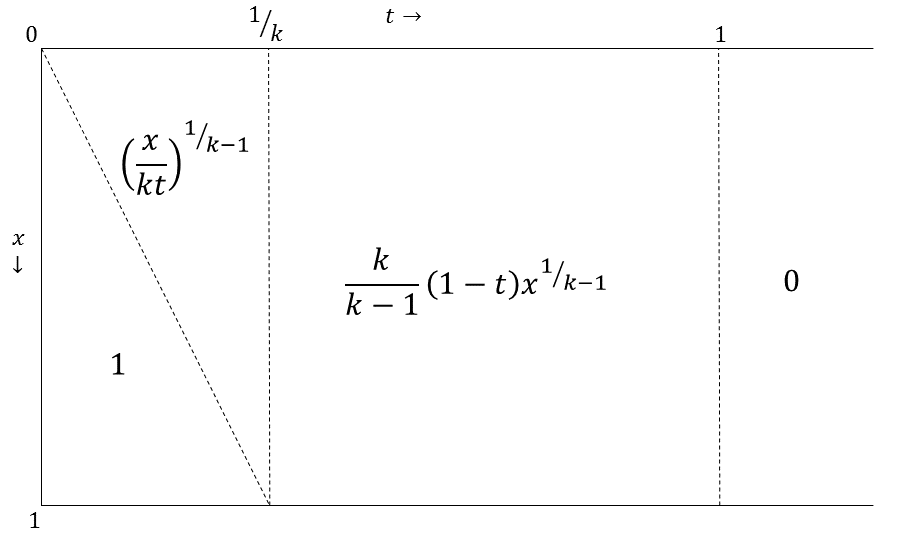}
\includegraphics[scale=0.3]{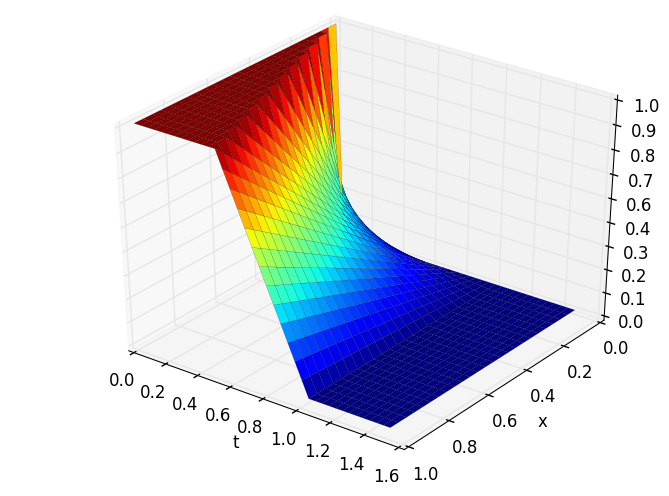}
\caption{The optimal function $\OPT{k}$.}
\label{fig:opt}
\end{figure}

For each $k \geq 2$ we define the function $\OPT{k} \in \cont((0,1])$:
\BNEQ \label{eq:opt}
\OPT{k}(x,t) = \begin{cases}
1 & t < x/k \\
\BF{x}{kt}^\R{k-1} &  x/k < t < 1/k\\
\F{k}{k-1}(1-t)x^\R{k-1} & 1/k < t < 1\\
0 & t > 1
\end{cases}
\ENEQ
See Figure \ref{fig:opt} for an illustration of this function.
\begin{claim} \label{clm:opt}
$\OPT{k} \in \valid((0,1])$,
 and every $N \in \valid((0,1])$ satisfies $\theta_N(k) \geq \theta_\OPT{k}(k)$.
Also,
\BEQ
\theta_\OPT{k}(k) = \F{3k -1}{k(k+1)}
\EEQ
\end{claim}
The proof appears in Appendix \ref{apx:opt}. It proceeds as suggested above: using Lemma \ref{lm:newLemma} we know how each column of the optimal function looks like, we then use the column requirements to find them exactly. Eventually we calculate the double integral and get $\theta_{\OPT{k}}(k)$.

\subsection{And now with $\epsilon$} \label{sct:epsilon}

The following claim is what we need to conclude the upper bound proof and prove Claim \ref{clm:upper}:
\begin{claim}
There is some real function $f$, s.t. $\lim_{\epsilon \rightarrow 0} f(\epsilon) = 1$, and every $N \in \valid([\epsilon, 1])$ satisfies
$
\theta_N(k) \geq  f(\epsilon) \cdot \theta_\OPT{k}(k) 
$.
\end{claim}
\BPF
We construct a new function $N'$ that will span the whole range of $x$'s from $0$ to $1$, with little change to $\theta_N(k)$.
This will be done by slowing $N$ down, and using what we saved in the column integrals to visit the $x$'s between $0$ and $\epsilon$ using our optimal solution, running it fast enough so it does not incur a big difference in $\theta_N(k)$.

Fix some $a < 1$ to be determined later. Define:
\BEQ
N'(x,t) = 
\begin{cases}
\zoom{\OPT{k}}{\epsilon}{\epsilon/(1-a)} & x \leq \epsilon \\
\zoom{N}{1}{1/a} & x > \epsilon
\end{cases}
\EEQ
Since the zoomed version of $\OPT{k}$ here is defined on the $x$'s in $(0,\epsilon]$ and the zoomed $N$ is on those in $[\epsilon, 1]$, we get that for all $t$:  
\BEQ
\column_{N'}(t) 
 = 
\column_{\zoom{\OPT{k}}{\epsilon}{\epsilon/(1-a)}}(t) + \column_{\zoom{N}{1}{1/a}}(t) 
= 
\epsilon \column_{\OPT{k}}\B{\F{1-a}{\epsilon} t} + \column_N(at)
 \leq
\epsilon \F{1-a}{\epsilon} t + at = t
\EEQ
where we used Claim \ref{clm:zoom} and the fact that both $\OPT{k}$ and $N$ satisfy the column requirements.

$N'$ can be made to be continuous with negligible cost to the integrals, as it is not continuous only on the seam between the two functions. Therefore, $N' \in \valid((0,1])$ and by Claim \ref{clm:opt}, we have $\theta_{N'}(k) \geq \theta_\OPT{k}(k)$.
So, again by Claim \ref{clm:zoom}:
\BEQ
\theta_\OPT{k}(k) 
 \leq
\theta_{N'}(k) 
= 
\epsilon \cdot \theta_{\zoom{\OPT{k}}{\epsilon}{\epsilon/(1-a)}}(k)
+
(1-\epsilon) \cdot \theta_{\zoom{N}{1}{1/a}}(k)
 =
\F{\epsilon}{1-a} \theta_\OPT{k}(k)+  \F{1-\epsilon}{a} \theta_N(k)
\EEQ
And therefore,
\BEQ
\F{1-\epsilon}{a}\theta_N(k) \geq \B{1 - \F{\epsilon}{1-a}} \theta_\OPT{k}(k) 
\quad \Longrightarrow \quad
\F{\theta_N(k)}{\theta_\OPT{k}(k)} \geq \F{a}{1-\epsilon}\B{1 - \F{\epsilon}{1-a}} 
\EEQ
Taking $a = 1 - \sqrt{\epsilon}$, the right size of the inequality becomes:
\BEQ
\F{1-\sqrt{\epsilon}}{1-\epsilon}\B{1 - \F{\epsilon}{\sqrt{\epsilon}}}
=
\F{\B{1-\sqrt{\epsilon}}^2}{1-\epsilon} = \F{{1-\sqrt{\epsilon}}}{1+\sqrt{\epsilon}}
\EEQ
As $\epsilon$ goes to $0$ this goes to 1.
\EPF

\section{Matching Lower Bound}

 In this section we prove:
\begin{theorem} \label{thm:lower}
For all $k \geq 1$:
\BEQ
\speedup(k) \geq \F{k(k+1)}{3k - 1}
\EEQ
\end{theorem}
The case of $k=1$ was dealt with in Section \ref{sct:one}.
For $k \geq 2$, denote by $A_{k,M}$, Algorithm \ref{alg:new} when set to run with parameter $k$ and $M$.
To prove Theorem \ref{thm:lower}, we prove:
\begin{claim} \label{clm:alg1}
For every $k \geq 2$, 
\BEQ
\lim_{M \rightarrow \infty} \theta_{A_{k,M}}(k) \leq \F{3k-1}{k(k+1)}
\EEQ
\end{claim}
\BPF
Denote the matrix of $A_{k,M}$ by $N$.
A general observation is that:
\BEQ
N(x,t) = N(x,t-1) \cdot \prob{\text{$x$ is not chosen at time $t$}}
\EEQ
Since the algorithm chooses uniformly from a set of unopened boxes at each stage, this is:
\BEQ
N(x,t) = N(x,t-1) \cdot \B{1 - \R{|\text{interval chosen from}| - (t-1) }}
\EEQ
Applying generously we get:
\BEQ
N(x,t) = \begin{cases}
1 & t < x/k \\
\prod_{i=\ceil{x/k}}^t \B{1 - \R{ik - i + 1}} & x/k \leq t \leq M/k \\
\prod_{i=\ceil{x/k}}^{M/k} \B{1 - \R{ik - i + 1}} 
\prod_{i=M/k+1}^t \cdot \B{1 - \R{M - i+1}} & M/k < t < M \\
0  & t \geq M
\end{cases}
\EEQ
where we assumed $M$ is a multiple of $k$. Otherwise, take $M$ large enough and round it up to be a multiple of $k$, the few extra boxes will not change the speed-up.
Also, note that the value of $N(x,t)$ where $x$ is not a multiple of $k$ is the same as that of $N(x',t)$ where $x'$ is the next multiple of $k$ greater than $x$. We can therefore calculate the double sum of $\theta_N(k)$, when summing only over $x$'s that are multiples of $k$, and multiply the result by $k$. From here on we will assume $x$ is a multiple of $k$.

We will need the following lemma:
\begin{lemma} \label{lm:gamma}
For integers $b \geq a \geq 1$, and $0 < \phi \leq 1$,
$
\prod_{i=a}^b \F{i}{i+\phi} \leq \BF{a}{b}^\phi
$.
\end{lemma}
Using properties of the Gamma function it is easy to see that the two sides of the equation are asymptotically equal, but this is not enough to prove our result as we need the inequality for small $a$ and $b$ as well. We prove this Lemma in Appendix \ref{apx:gamma}. Now:
\BEQ
\prod_{i=\F{x}{k}}^t \B{1 - \R{ik - i + 1}}
=
\prod_{i=\F{x}{k}}^t \F{ik - i}{ik -i + 1}
=
\prod_{i=\F{x}{k}}^t \F{i}{i + \R{k-1}}
\leq
\B{\F{x}{kt}}^\R{k-1}
\EEQ
And:
\BEQ
\prod_{i=\F{M}{k}+1}^t \B{1 - \R{M - i+1}}
 & = 
\prod_{i=\F{M}{k}+1}^t \F{M-i}{M-i+1}
 =
\F{M-\F{M}{k}-1}{M-\F{M}{k}}
\cdots
\F{M-t}{M-t+1}
\\ &  = 
\F{M-t}{M-\F{M}{k}} = \F{1-\F{t}{M}}{1-\R{k}} = \F{k}{k-1}\B{1 - \F{t}{M}}
\EEQ
So finally:
\BEQ
N(x,t) \leq \begin{cases}
1 & t < x/k \\
\BF{x}{kt}^\R{k-1} & x/k \leq t \leq M/k \\
\F{k}{k-1}\B{1 - \F{t}{M}} \BF{x}{M}^\R{k-1}  & M/k < t < M \\
0 & t \geq M
\end{cases}
\EEQ
This is strangely familiar. Looking at \eqref{eq:opt} we notice that: 
\BEQ
N(x,t) \leq \OPT{k}(x/M, t/M)
\EEQ
Calculating the sums of $\theta_N(k)$ is like approximating the double integral of 
$\theta_{\OPT{k}}$, and as $M$ goes to infinity this approximation can be as close as we wish. A fine detail is with the fact that we take only $x$'s that are multiples of $k$, but
this creates a $1/k$ factor in the sum approximating the integral which is then cancelled by multiplying the total sum by $k$, as said above.
\EPF

\subsection{Algorithm with ${O(\log(M))}$ memory}

As it is now, Algorithm \ref{alg:new} uses a lot of memory. Each searcher needs to remember all of the boxes it checked, so we need $O(M)$ memory in the worst case. 
However, if we are willing to lose a little in the speed-up, we can save a lot in memory. 
\begin{algorithm}
\caption{A memory efficient variant.}\label{alg:mem}
\begin{algorithmic} 
\For{$i=1$ to $M/k$}
   \State check a uniformly chosen box in $\set{1, \ldots, i*k}$
\EndFor
\While{there are unchecked boxes}
	\State check a uniformly chosen box out of all boxes.
\EndWhile
\end{algorithmic}
\end{algorithm}
Algorithm \ref{alg:mem} is exactly the same as Algorithm \ref{alg:new}, except it doesn't remember where it's been. The only memory it uses is the counter, and so it needs only $O(\log(M))$ bits of memory. 
Denote by $B_{k,M}$, Algorithm \ref{alg:mem} when set to run with parameter $k$ and $M$.
In Appendix \ref{apx:smallMemory} we prove:
\begin{claim} \label{clm:logMemory}
$
\lim_{M \rightarrow \infty} \speedup_{B_{k,M}}(k) \geq \F{k}{3}
$
\end{claim}
As we can see, this algorithm, for large $M$ and $k$ will have essentially the same speed-up as that of the optimal algorithm while using exponentially less memory.

\appendix

\section{Appendix}

\subsection{Proof of Claim \ref{clm:Mat2Func}} \label{apx:matToFunc}

\newtheorem*{claim:mat2Func}{Claim \ref{clm:Mat2Func}}
\begin{claim:mat2Func}
For any algorithm $A$ that works on $M$ boxes, there is a function $N \in \valid([1,M+1])$ such that for all $k$,
$\theta_N(k) \leq \theta_A(k)$.
\end{claim:mat2Func}
\BPF
Given $A$, take its matrix $N_A$, and define
$N(x,t) = N_A(\lfloor x \rfloor ,\lfloor t \rfloor)$.
For any $t$:
\BEQ
\column_N(t) & = \int_1^{M+1} 1 - N(x,t) dx
=
\int_1^{M+1} 1 - N_A(\lfloor x \rfloor, \lfloor t \rfloor) dx
\\ & =
\sum_{x=1}^M 1 - N_A(x, \lfloor t \rfloor)
= \column_{N_A}(\lfloor t \rfloor) \leq 
\lfloor t \rfloor \leq t
\EEQ

So $N$ satisfies the column requirements. Next, let us upper-bound the inverse of the speed-up.
\BEQ
\theta_N(k) 
& = 
\R{M} \int_1^{M+1} \int_0^\infty \R{x} N_A(\lfloor x \rfloor ,\lfloor t \rfloor)^k dt dx
\leq
\R{M} \int_1^{M+1} \int_0^\infty \R{\lfloor x \rfloor} N_A(\lfloor x \rfloor ,\lfloor t \rfloor)^k dt dx
\\ & =
\R{M} \sum_{x=1}^M \sum_{t=0}^\infty \R{x} N_A(x, t) = \theta_A(k)
\EEQ
where the inequality is because $x \geq 1$ and $\lfloor x \rfloor \leq x$. 
$N$ is a step function and not continues but can be turned into a continues function with as little change in integral as we wish - we can leverage what we gain in the inequality above to account for this change.
\EPF

\subsection{Proof of the Extended Lemma} \label{apx:mainLemma}

\newtheorem*{lemma:extended}{Lemma \ref{lm:newLemma}}
\begin{lemma:extended}
Fix $T \geq 0$.
For continuous functions $a: (0,1] \rightarrow \Real^+$ and $f : (0,1] \rightarrow [0,1]$ where $\int_0^1 a(x)^{-1/(k-1)} dx$ is defined and $\int_0^1 1 - f(x) dx \leq T$, 
the minimum of $\int_0^1 a(x) f(x)^k dx$, ranging over all possible $f$'s is achieved when:
\BEQ
f(x) = \min\B{1, \alpha a(x)^{-\R{k-1}}}
\EEQ
Where $\alpha$ is a function of $a(\cdot)$ and $T$, and independent of $x$. 
\end{lemma:extended}

\BPF
First, note that if $T \geq 1$ then the optimal solution is $f(x) = 0$ and then $\alpha = 0$. Otherwise, and this hold for Lemma \ref{lm:oldLemma} as well, we can assume that the constraint is actually an equality, as maximizing it will always decrease the target integral.

For each $0 < s < 1/2$, applying Lemma \ref{lm:oldLemma},  there is some $\alpha_s$ such that among all functions $f:[s,1] \rightarrow [0,1]$, satisfying $\int_s^1 1 - f(x)dx \leq T - s$ (again, since $T < 1$, this is actually an equality), the function:
\BEQ
f_s(x) = \min(1, \alpha_s a(x)^{-1/(k-1)})
\EEQ
Minimizes $\int_s^1 a(x) f(x)^k dx$.

Extend each $f_s$ to be also defined on $(0,s)$ by setting it to be $0$ there.
Since $\alpha_s$ is derived as the solution to:
\BEQ
\int_s^1 1 - \min(1, \alpha_s a(x)^{-\R{k-1}}) dx = T - s
\quad \Longrightarrow \quad
\int_s^1 \min(1, \alpha_s a(x)^{-\R{k-1}}) dx = 1 - T
\EEQ
We can see that as $s$ decreases, $\alpha_s$ decreases, and as the $\alpha_s$ are bounded below by $0$, they converge to some $\alpha$. Denote $f(x) = \min(1, \alpha a(x)^{-1/(k-1)})$.

Clearly, the $f_s$ converge point-wise to $f$ as $s$ approaches $0$. By the bounded convergence theorem we get:
\BEQ
\int_0^1 f(x)dx 
= \lim_{s\rightarrow 0} \int_0^1 f_s(x)dx 
= \lim_{s\rightarrow 0} \B{\int_0^s 0 dx + \int_s^1 f_s(x)dx}
= T
\EEQ
Since $\alpha_s$ decreases as $s$ decreases, we get that for all $s < 1/2$:
\BEQ
a(x) f_s^k(x) = 
a(x) \min\B{1, \alpha_s a(x)^{-\R{k-1}}}^k \leq
a(x) \B{\alpha_s a(x)^{-\R{k-1}}}^k \leq
 \alpha^k_\R{2} a(x)^{-\R{k-1}} 
\EEQ
By our requirements on $a(\cdot)$ we know that $\int_0^1 a(x)^{-\R{k-1}}dx < \infty$, so by the dominated convergence theorem we can exchange integral and limit and get:
\BNEQ \label{eq:lmUniform}
\int_0^1 a(x) f(x)^k dx =
\lim_{s\rightarrow 0} \int_0^1 a(x) f_s(x)^k dx
\ENEQ 
Now, assume there is some $g$ satisfying the conditions that is better than $f$. That is:
\BNEQ \label{eq:lmOneSide}
\int a(x) f(x)^k - \int a(x) g(x)^k > \delta > 0
\ENEQ
We take an $s$ small enough so that \eqref{eq:lmUniform} is more than $\delta/2$ close to the limit, and we get:
\BEQ
\int_0^1 a(x) f(x)^k - \int_0^1 a(x) g(x)^k < 
\F{\delta}{2} + \int_0^1 a(x) f_s(x)^k - \int_0^1 a(x) g(x)^k
\EEQ
Yet we know that $f_s$ is optimal on $[s,1]$ and since it is $0$ on $(0,s)$, $g$ cannot be better. so this latter sum is at least $\delta/2$ contradicting \eqref{eq:lmOneSide}.
\EPF

\subsection{Optimality of $\OPT{k}$} \label{apx:opt}

\newtheorem*{claim:opt}{Claim \ref{clm:opt}}
\begin{claim:opt} 
$\OPT{k} \in \valid((0,1])$,
 and any  $N \in \valid((0,1])$ has $\theta_N(k) \geq \theta_\OPT{k}(k)$.
Also,
\BEQ
\theta_\OPT{k}(k) = \F{3k -1}{k(k+1)}
\EEQ
\end{claim:opt}
\BPF
Corollary \ref{cr:main} gives for each fixed $t$ the function $f_t(x)$ minimizing
$\int_0^1 \R{x} f(x)^k dx$. Setting $\OPT{k}(x,t) = f_t(x)$ will give the minimal
$\int_0^\infty \int_0^1 \R{x} \OPT{k}(x,t)^k dx dt$, which is the goal.

All we have to do is figure out $f_t(x)$. 
First, for $t > 1$, the optimal $f_t$ is obviously $f_t(x) = 0$, since this satisfies the column requirement and contributes $0$ to our target double integral.
For $t<1$, we have $\int_0^1 1 - f_t(x)dx \leq t$,
and that $f_t(x) = \min(1, \alpha x^{1/(k-1)})$.
So, given $t$, we figure out what its $\alpha$ is.
For each $t$, denote by $\gamma$ (a function of $t$) the smallest $x$ where $N(x,t) = 1$, and in case this does not happen, set $\gamma = 1$:
\BNEQ \label{eq:gamma}
\gamma = \min(1, \R{\alpha^{k-1}})
\ENEQ
for every $t<1$, $\OPT{k}$ satisfies the column requirement $\column_\OPT{k}(t) \leq t$, and since it minimizes the target integral, it will actually be an equality:
\BEQ
1 - t = \int_0^1 \OPT{k}(x,t) dx
\EEQ
We have two cases:
\BE
\I
for all $t$ where $\gamma < 1$ this equation is:
\BEQ
1 - t = \int_0^\gamma \alpha x^\R{k-1}  + \int_\gamma^1 1 dx 
\quad \Longrightarrow \quad
\gamma - t = \int_0^\gamma \alpha x^\R{k-1} dx
\EEQ
From \eqref{eq:gamma}, we get $\alpha = 1/\gamma^{1/(k-1)}$. Plugging this in:
\BEQ
\gamma -t =
\int_0^\gamma \left(\F{x}{\gamma}\right)^\R{k-1} dx 
= \gamma \int_0^1 x^\R{k-1} dx = \gamma \R{\R{k-1} + 1} = 
\F{k-1}{k} \gamma
\EEQ
So:
\BEQ
\gamma - t = \gamma \F{k-1}{k} \quad \Longrightarrow \quad 
\gamma = kt
\EEQ
This means, that for all $t < \R{k}$, $\gamma < 1$ and then 
$\alpha = 1/(kt)^{1/(k-1)}$. For all other $t$, $\gamma = 1$.
\I
for all $t$ where $\gamma = 1$:
\BEQ
1 - t 
= \int_0^1 \alpha x^\R{k-1} dx
= \alpha \R{\R{k-1} + 1} = \F{k-1}{k} \alpha 
\EEQ
and so:
\BEQ
\alpha = \F{k}{k-1} (1 - t)
\EEQ
\EE
Putting all this together gives us $\OPT{k}(x,t)$ as in \eqref{eq:opt}.

Now we can calculate $\theta_\OPT{k}(k)$:
\BEQ
\int_0^1 \int_0^1 \R{x} \OPT{k}(x,t)^k dx dt 
& = 
\int_0^{1/k} \left( \int_0^\gamma \R{x}(\alpha x^\R{k-1})^k dx \right) dt
+    
\int_0^{1/k} \left( \int_\gamma^1 \R{x}1^k dx \right) dt \\
&  +
\int_{1/k}^1 \left(  \int_0^1 \R{x}(\alpha x^\R{k-1})^k dx  \right) dt
\EEQ
We focus on each one of these:
\BE

\I
Here we know that $\gamma = kt < 1$.
\BEQ
\int_0^\gamma \R{x}(\alpha x^\R{k-1})^k dx   
= \alpha^{k-1} \int_0^\gamma \alpha x^\R{k-1} dx  
= \alpha^{k-1} ( \gamma - t) = \R{\gamma} (\gamma - t)
\EEQ
We get: 
\BEQ
\R{kt} (kt - t) = \F{k-1}{k}
\EEQ
The whole integral:
\BEQ
\int_0^{1/k} \F{k-1}{k} = \F{k-1}{k^2}
\EEQ

\I
Here, still, $\gamma < 1$.
\BEQ
\int_\gamma^1 \R{x} dx = \log(1) - \log(\gamma) = -\log(\gamma) 
\EEQ
Plugging our $\gamma = kt$ in, and calculating the whole integral:
\BEQ
\int_0^{1/k} - \log(kt) dt = - \R{k} \int_0^1 \log(t)dt  = \R{k}
\EEQ
Last bit is because indefinite integral of $\log(x)$ is $x\log(x) - x$.
\I
Here $\gamma = 1$.
\BEQ
\int_0^1 \R{x}(\alpha x^\R{k-1})^k dx
= 
\alpha^{k-1} \int_0^1 \alpha x^\R{k-1} dx  
= 
\alpha^{k-1} (1 - t)
\EEQ
plugging in our $\alpha = \F{k}{k-1}(1-t)$ and calculating the whole integral:
\BEQ
&
\int_{1/k}^1 \left(\F{k}{k-1} (1 - t)\right)^{k-1} (1 - t) dt 
=
\left(\F{k}{k-1}\right)^{k-1} \int_0^{\F{k-1}{k}} t^k dt 
\\ & =
\left(\F{k}{k-1}\right)^{k-1} \R{k+1} \left( \F{k-1}{k} \right)^{k+1}
 =
\F{(k-1)^2}{k^2(k+1)}
\EEQ 
\EE 
In total we get:
\BEQ
&
\F{k-1}{k^2} + \R{k} + \F{(k-1)^2}{k^2(k+1)}
 = 
\R{k} \cdot
\F{(k-1)(k+1) + k(k+1) + (k-1)^2}{k(k+1)}
\\ &= 
\R{k} \cdot
\F{k^2 - 1 + k^2+k + k^2-2k+1}{k(k+1)}
= 
\R{k} \cdot
\F{3k^2-k}{k(k+1)}
= \F{3k-1}{k(k+1)}
\EEQ

\EPF

\subsection{Proof of the Gamma Function Property} \label{apx:gamma}

\newtheorem*{lemma:gamma}{Lemma \ref{lm:gamma}}
\begin{lemma:gamma}
For integers $b \geq a \geq 1$, and $0 < \phi \leq 1$,
\BEQ
\prod_{i=a}^b \F{i}{i+\phi} \leq \BF{a}{b}^\phi
\EEQ
\end{lemma:gamma}
\BPF
By induction on $a$ (somehow on $b$ it doesn't work..).
If $b = a$, then we should show 
$a/(a + \phi) \leq 1$, which is true.
We therefore assume that the result holds for $a+1$ and prove it for $a$:
\BEQ
\prod_{i=a}^b \F{i}{i+\phi}
= \F{a}{a+\phi} \cdot \prod_{i=a+1}^b \F{i}{i+\phi} 
\leq \F{a}{a+\phi} \cdot \BF{a+1}{b}^\phi
\EEQ 
We want to show:
\BEQ
& \F{a}{a+\phi} \cdot \BF{a+1}{b}^\phi  \leq \BF{a}{b}^\phi 
\quad \Longleftrightarrow \quad
\F{a}{a+\phi}  \leq \BF{a}{a+1}^\phi  
\\ & 
\quad \Longleftrightarrow \quad
\B{\F{a+\phi}{a}}^\R{\phi}  \geq \B{\F{a+1}{a}}
\EEQ
Take $b=\R{a}<1$ and $x=\R{\phi}\geq 1$ the above is equivalent to:
\BEQ
\B{1+\F{b}{x}}^x & \geq \B{1+b}
\EEQ
If we show that the left side is increasing with $x$ when $x\geq 1$ then we are done.
We take the derivative (using an internet site):
\[
\B{1+\F{b}{x}}^x \cdot \left( \ln\B{1+\F{b}{x}}  - \F{b}{x\B{1+\F{b}{x}}}\right)
\]
This is positive if
\[
\B{1+\F{b}{x}}\ln\B{1+\F{b}{x}}  > \F{b}{x}
\]
Take $y=\F{b}{x}\leq 1$. We want to show that 
\[
\B{1+y}\ln\B{1+y}  > y
\]
We use the equality
\[
\ln\B{1+y} = \int_0^y \R{1+t}dt
\]
So 
\[
\B{1+y}\ln\B{1+y} = \int_0^y \F{1+y}{1+t}dt> \int_0^y  1 dt =y
\]
 as desired.
\EPF

\subsection{Speedup of Algorithm \ref{alg:mem}} \label{apx:smallMemory}

\newtheorem*{claim:logMemory}{Claim \ref{clm:logMemory}}
\begin{claim:logMemory}
\BEQ
\lim_{M \rightarrow \infty} \speedup_{B_{k,M}}(k) \geq \F{k}{3}
\EEQ
\end{claim:logMemory}
\BPF
By the same reasoning as the original algorithm, the matrix of this one is:
\BEQ
N(x,t) = \begin{cases}
1 & t < x/k \\
\prod_{i=\ceil{x/k}}^t \B{1- \R{ik}} &  x/k \leq t \leq M/k \\
\prod_{i=\ceil{x/k}}^{M/k} \B{1- \R{ik}}\cdot \B{1-\R{M}}^{t-M/k} &   t > M/k
\end{cases}
\EEQ
Using Lemma \ref{lm:gamma}:
\BEQ
\prod_{i=x/k}^t \B{1 - \R{ik}} \leq 
\prod_{i=x/k}^t \B{1 - \R{ik+1}} =
\prod_{i=x/k}^t \F{ik}{ik+1} =
\prod_{i=x/k}^t \F{i}{i + \R{k}} \leq \BF{x}{tk}^\R{k}
\EEQ
Using it and ignoring all the rounding, as in the analysis of Algorithm \ref{alg:new}:
\BEQ
N(x,t) \leq \begin{cases}
1 & t < x/k \\
\BF{x}{tk}^\R{k}  &  x/k \leq t \leq M/k \\
\BF{x}{M}^\R{k} \cdot \B{1-\R{M}}^{t-M/k} &   t > M/k
\end{cases}
\EEQ
To upper bound $\theta_N(k) = \R{M} \sum_{x=1}^M \R{x} \sum_{t=0}^\infty N(x,t)^k$, we calculate each area of the sum separately. 
In what follows, we use the fact that $M$ is large for a few approximations.
\BE
\I
The $1$'s contribute:
\BEQ
\R{M} \sum_{x=1}^M  \R{x} \sum_{t=0}^{x/k} 1^k =
\R{M} \sum_{x=1}^M  \R{k} = \R{k}
\EEQ
\I
The second part contributes:
\BEQ
&
\R{M} \sum_{x=1}^M  \R{x} \sum_{t=x/k}^{M/k} \F{x}{tk}
=
\R{M} \sum_{x=1}^M  \sum_{t=x/k}^{M/k} \R{tk}
\approx
\R{Mk} \sum_{x=1}^M  \sum_{t=x}^{M} \R{t}
=
\R{Mk} \sum_{x=1}^M  \sum_{t=x}^{M} \R{t}
= \R{Mk} M = \R{k} 
\EEQ
where the double sum is equal to $M$, by noting that $1$ appears there once, $\R{2}$ appear twice, and so on.
\I
The last part is:
\BEQ
&
\R{M} \sum_{x=1}^M  \R{x} \BF{x}{M} 
\sum_{t=M/k}^\infty \B{1-\R{M}}^{\B{t - \F{M}{k}}k}
=
\R{M^2} \sum_{x=1}^M  \sum_{t=0}^\infty \B{1-\R{M}}^{tk}
\\ & =
\R{M} \R{1-\B{1-\R{M}}^k}
\approx
\R{M} \R{1-\B{1-\F{k}M}} = \R{k}
\EEQ
\EE
Putting it all together, we get the result.
\EPF

\bibliography{bib}

\end{document}